\documentstyle[prb,aps,epsf]{revtex}
\newcommand{\ti}[1]{{\mbox{\scriptsize #1}}}
\newcommand{\thefigurename}{Figure}  
\def\fnum@figure{\thefigurename\ \thefigure}
\renewcommand{\thefigurename}{Fig.}
\begin{document}
\draft
\title{Instanton calculation of the density of states of
disordered Peierls chains}
\author{Maxim Mostovoy\cite{Perm} and Jasper Knoester}
\address{Institute for Theoretical Physics
and Materials Science Center\\
University of Groningen, Nijenborgh 4, 9747 AG Groningen, The Netherlands}
\date{\today}
\maketitle

\begin{abstract}

We use the optimal fluctuation method to find the density of
electron states inside the pseudogap in disordered Peierls
chains.  The electrons are described by the one-dimensional Dirac
Hamiltonian with randomly varying mass (the Fluctuating Gap
Model).  We establish a relation between the disorder average in
this model and the quantum-mechanical average for a certain
double-well problem.  We show that the optimal disorder
fluctuation, which has the form of a soliton-antisoliton pair,
corresponds to the instanton trajectory in the double-well
problem.  We use the instanton method developed for the
double-well problem to find the contribution to the density of
states from disorder realizations close to the optimal
fluctuation.

\end{abstract}

\section{Introduction}

In studies of disordered systems, one often encounters the
problem of finding the density of single-particle states in
energy regions where the states can only exist as a consequence
of disorder.  An example is the density of negative energy states
of a particle moving in a random potential (the lowest energy of
a free particle is zero).\cite{HL,ZL,Ltail} Another example is
provided by a superconductor with magnetic impurities.  The
impurities break Cooper pairs and fill the gap with electron
states of arbitrarily small energy.\cite{BT} The density of such
states (``Lifshitz tails'') is typically small, as it requires a
large disorder fluctuation to induce them.  In that case the
density can be estimated by the probability of the least
suppressed disorder fluctuation that creates a state of a given
energy (the so-called optimal fluctuation
method).\cite{HL,ZL,Ltail,BT}

In Ref.~\CITE{MKPL} we applied the optimal fluctuation method to
study the density of states and optical absorption in disordered
Peierls insulators.  In the absence of disorder, the electron
excitation spectrum in these quasi-one-dimensional materials has
a gap which results from the electron-lattice instability and is
accompanied by a periodic lattice distortion with the wave vector $Q
= 2 k_F$.\cite{Peierls} We considered half-filled chains, in
which case the Peierls instability results in an alternation of
long and short bonds (dimerization). Disorder in the
electron hopping amplitudes induces electron states inside the
Peierls gap, transforming it into a pseudogap. In
Ref.~\CITE{MKPL} we showed that the optimal fluctuation for a
disordered dimerized chain has the form of a soliton-antisoliton
pair.  It induces two electron states which lie close to the
center of the pseudogap and which are symmetric and antisymmetric
superpositions of states localized near the soliton and
antisoliton.  Knowing the typical form of the disorder-induced
electron states we were able to find the absorption coefficient
at small energies. These results were also used to explain the
coexistence of the dimerization and antiferromagnetism, observed
recently in the spin-Peierls material CuGeO$_3$.\cite{MKK}

The probability of the optimal configuration gives, however, only
an estimate of the density of states.  A more careful calculation
requires finding the contribution of disorder realizations close
to the optimal one.  In this paper, we perform such a calculation
using the close relation between the optimal fluctuation method
and the semiclassical approximation in quantum mechanics and
field theory.  In the path integral version of this
approximation, the most important paths lie close to the
``saddle-points'' of the action, called
instantons.\cite{Polyakov,Coleman} In particular, the instantons
for the double-well problem are the classical trajectories
describing the imaginary time motion between the two wells, which
corresponds to tunneling and results in level splitting.  We
show how the disorder average of the density of states can be
written in the form of a functional integral that represents the
sum over all paths of a particle in a corresponding double-well
potential.  Then the optimal disorder configuration becomes
precisely the instanton trajectory and the calculation of the
density of states can be done using standard instanton methods.

The outline of this paper is as follows.  In Sec.~\ref{FGM} we
introduce the Fluctuating Gap Model describing disordered Peierls
systems and briefly discuss the previously obtained analytical
results for the disorder-averaged density of states in that
model.  Next, in Sec.~\ref{funint} we establish the relation
between the disorder average in the Fluctuating Gap Model and the
quantum-mechanical average for a certain double-well problem.  We
find the instanton trajectory corresponding to the optimal
fluctuation and perform the ``saddle-point'' integration for the
average density of states.  We discuss our results and conclude
in Sec.~\ref{discussion}.  Two technical points are addressed in
appendices.  In Appendix A we discuss how the symmetries of the
FGM Hamilitonian are crucial for the validity of the optimal
fluctuation method.  In Appendix B we discuss the instanton
calculation for a different choice of boundary conditions.

\section{The Fluctuating Gap Model}
\label{FGM}

The Fluctuating Gap Model (FGM) is a continuum model describing
electrons in disordered quasi-one-dimensional
semiconductors.\cite{Keldysh,LSP} In this paper we use it to
describe half-filled Peierls chains with disorder in the hopping
amplitudes of the electrons between the chain's lattice sites.
These systems are characterized by the Peierls order parameter,
$\Delta(x)$, which is the (continuum version of the) alternating
part of the hopping amplitudes along the chain, {\em i.e.}, the
difference between the hopping amplitudes on nearest odd and even
bonds.  In the absence of disorder this order parameter (also
called ``dimerization'') is constant along the chain and equals
the gap in the single-electron spectrum.\cite{Peierls,SSH}
Disorder in the electron hopping amplitudes results in random
variations of the order parameter along the chain (which explains
the name of the model):
\begin{equation}
\Delta(x) = \Delta_0 + \eta(x) \;.
\label{Delta(x)}
\end{equation}
Here, $\Delta_0$ is the average value of the order parmeter and
$\eta(x)$ is the fluctuating part with a Gaussian correlator,
\begin{equation}
\label{Gauss}
\langle \eta (x) \eta(y) \rangle = A \delta(x - y) \;.
\end{equation}

The single-electron states close to the Fermi energy
$\varepsilon_F = 0$, are described by the wave function
\[
\psi(x) = \left( \begin{array}{c}
\psi_{1}(x) \\
\psi_{2}(x) \\
\end{array} \right) \;,
\]
where the two amplitudes $\psi_{1}(x)$ and $\psi_{2}(x)$
correspond to electrons moving, respectively, to the right and to
the left with the Fermi velocity $v_F$.  Since the density of
single-electron states does not depend on spin projection, we
omit the spin index of the electron wave function.  The wave
functions satisfy the one-dimensional Dirac equation,
\begin{equation}
\label{Dirac}
{\hat h} \psi = \left( \sigma_3 \frac{v_F}{i} \frac{d}{dx} +
\sigma_1 \Delta(x) \right) \psi(x) = \varepsilon \psi(x) \;,
\end{equation}
where $\sigma_1$ and $\sigma_3$ the Pauli matrices.  For the
applicability of the continuum description of electrons, the
random variations in the hopping amplitudes have to be relatively
small.

This model was first introduced by Keldysh.\cite{Keldysh} It was
also considered in the context of the thermodynamical properties
of quasi-one-dimensional charge-transfer salts,\cite{OE} and has
been applied to study the effect of disorder on the Peierls
transition temperature,\cite{XT,H&F} as well as the effect of
quantum lattice fluctuations on the optical spectrum of Peierls
materials.\cite{Kim,H&M,S&M} Recently it was used to describe the
phase diagram of disordered spin-Peierls systems.\cite{MKK}

In the remainder of this section we discuss the dependence of the
spectrum of the single-electron eigenstates on the disorder
strength. By $\rho(\varepsilon)$ we denote the disorder averaged
density of states per unit length:
\begin{equation}
\rho(\varepsilon) = \frac{1}{L}
\langle \mbox {Tr}
\left[ \delta({\hat h} - \varepsilon) \right] \rangle\;,
\end{equation}
where $L$ is the chain length.  Due to the charge conjugation
symmetry of the Dirac Hamiltonian, the density of states is a
symmetric function of energy:
\begin{equation}
\rho(-\varepsilon) = \rho(\varepsilon)\;.
\label{symrho}
\end{equation}
We also introduce $N(\varepsilon)$, equal to the average number
of electronic states (per unit length) with energy
between $0$ and some $\varepsilon > 0$:
\begin{equation}
N(\varepsilon) = \int_0^\varepsilon\!\!d\varepsilon^{\prime}
\rho(\varepsilon^{\prime})\;.
\label{N(E)}
\end{equation}
An analytical expression for $N(\varepsilon)$ was found by
Ovchinnikov and Erikhman\cite{OE} by means of the ``phase
formalism'',\cite{LSP,Schmidt}
\begin{equation}
N(\varepsilon) = \frac{2A}{(\pi v_F)^2}
\frac{1}{\left(J_{\nu}^2(z) + N_{\nu}^2(z)\right)}\;.
\end{equation}
Here, $z  =  v_F \varepsilon / A$ and $\nu  =  v_F \Delta_0 / A$
are dimensionless variables, and $J_{\nu}(z)$ and $N_{\nu}(z)$
are, respectively, Bessel and Neumann functions.

\begin{figure}[htbp]
\centering \leavevmode
\epsfxsize=7cm \epsfbox{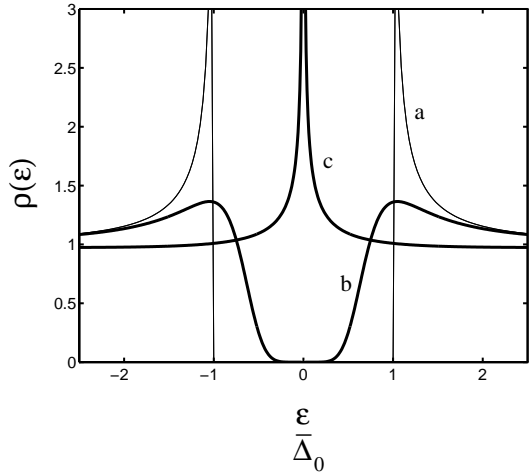}
\caption{ \footnotesize
Disorder-averaged density of states $\rho(\varepsilon)$ in the
FGM for three values of the dimensionless disorder strength: $g =
0$, {\em i.e.}, no disorder (curve a), $g = 0.25$ (curve b), $g =
4$ (curve c).  The free electron density of states was set to
unity.
\label{D.O.S.}}
\end{figure}

In Fig.\ref{D.O.S.} we plot the average density of states
$\rho(\varepsilon)$ for several values of the dimensionless
disorder strength $g = 1 / \nu = A / (v_F \Delta_0)$.  In the
absence of disorder ($\Delta(x) = \Delta_0$), the electron
spectrum has a gap between the energies $\varepsilon=-\Delta_0$
and $\varepsilon=+\Delta_0$.  Disorder gives rise to the
appearance of electron states inside the gap.  The energy
dependence of the average density of states close to the Fermi
energy ($|\varepsilon| \ll \Delta_{0}$) is given approximately
by\cite{OE}
\begin{equation}
\rho(\varepsilon)  = \frac{2}{v_F g
\Gamma^2(\frac{1}{g})}
\left( \frac{\varepsilon}{2 g \Delta_0}
\right)^{\frac{2}{g} - 1}
\label{dos}
\end{equation}
(here and below we assume $\varepsilon$ to be positive, which is
sufficient in view of Eq.(\ref{symrho})).  For $g < 2$ the
density of states has a pseudogap (the Peierls gap filled with
disorder-induced states).  For $g > 2$ the pseudogap disappears
and the density of states becomes divergent at $\varepsilon = 0$.
This is a Dyson-type singularity,\cite{Dyson} which occurs in the
band center for random Hamiltonians with charge conjugation
symmetry.

\begin{figure}[htbp]
\centering \leavevmode
\epsfxsize=10cm \epsfbox{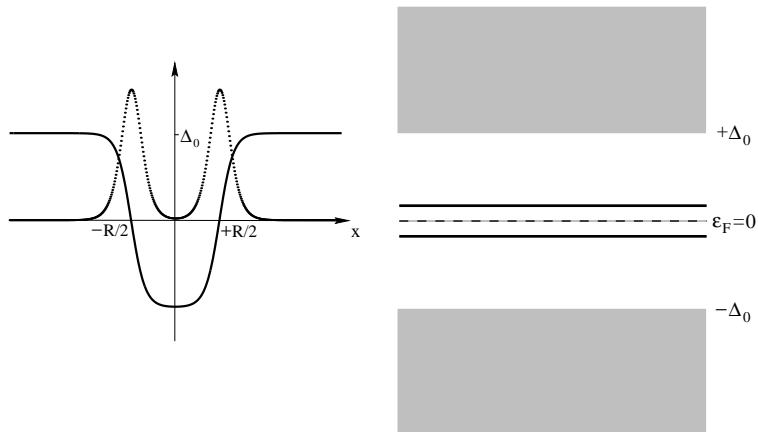}
\caption{ \footnotesize
The left part of the picture shows $\Delta(x) = \Delta_0 +
\bar{\eta}(x)$ for the instanton disorder fluctuation (thick
line) and the electron density $|\psi_+(x)|^2 = |\psi_-(x)|^2$
for the corresponding intragap states (dotted line); the right
part of the picture shows the spectrum of electron states for
the optimal disorder fluctuation.
\label{s-a}}
\end{figure}

In Ref.~\CITE{MKPL} we showed that the energy-dependence of the
average density of states at small energy and weak disorder is
mainly determined by the weight of the optimal disorder
fluctuation $\bar{\eta}(x)$, which is defined as the most
probable among the fluctuations that induce a state at given
energy $\varepsilon$.  The form of the optimal fluctuation is
\begin{equation}
\bar{\eta}(x) = - v_F K
\left[
\tanh\left( K(x - x_0 + \frac{R}{2}) \right) -
\tanh\left( K(x - x_0 - \frac{R}{2}) \right)
\right] \;,
\label{Soan}
\end{equation}
where $x_0$ and $R$ describe, respectively, the position and the
spatial extent of the disorder fluctuation, and $K$ is determined
by \begin{equation}
v_F K = \Delta_0 \tanh (KR)\;.
\label{VFK1}
\end{equation}
Thus, $\Delta(x) = \Delta_0 + \bar{\eta}(x)$, plotted in
Fig.~\ref{s-a}, has precisely the form of a soliton-antisoliton
pair discussed in Refs.~\CITE{BK,CB} in the context of polaron
excitations in the conjugated polymer {\em trans}-polyacetylene.
The spectrum of electron states for this $\Delta(x)$ (also
plotted in Fig.~\ref{s-a}) consists of a valence band (with
highest energy $- \Delta_{0}$), a conduction band (with lowest
energy $+\Delta_{0}$), and two localized intragap states
$\psi_{\pm}(x)$ with energies $\pm \varepsilon_0(R)$, where
\begin{equation}
\varepsilon_0(R) =  \frac{\Delta_{0}}{\cosh (KR)}\;.
\label{EPS}
\end{equation}
The soliton-antisoliton separation $R$ is fixed by the condition
$\varepsilon_0(R) = \varepsilon$.  The two intragap states are
the symmetric and antisymmetric superpositions of the midgap states
localized near the soliton and antisoliton.\cite{SSH} The energy
splitting $2 \varepsilon$ decreases exponentially with the
soliton-antisoliton separation, so that for $\varepsilon \ll
\Delta_0$,
\begin{equation}
R \approx  \xi_0 \ln \frac{2 \Delta_0}{\varepsilon} \;,
\label{dist}
\end{equation}
where $\xi_0 = v_F / \Delta_0$ is the correlation length.

For white-noise disorder with the correlator Eq.(\ref{Gauss}),
the weight $p\left[ \eta (x) \right]$ of the disorder
configuration $\eta(x)$ is given by
\begin{equation}
p\left[ \eta (x) \right] =
\exp \left( -\frac{1}{2A} \int\!\!dx\eta^{2}(x) \right) \;.
\end{equation}
Using Eq.(\ref{dist}) we find that the weight of the optimal
configuration,
\begin{equation}
p\left[ \bar{\eta}(x) \right] \propto
\varepsilon^{\frac{2}{g}}\;,
\label{weight}
\end{equation}
gives, indeed, a good estimate for the shape of the
density of states inside the pseudogap for $g \ll 1$ [cf.
Eq.(\ref{dos})].

We would like to conclude this section by noting that, while in
the random potential problem the optimal fluctuation induces one
state with large negative energy,\cite{HL,ZL,Ltail} in the FGM
the soliton-antisoliton fluctuation Eq.(\ref{Soan}) induces two
states with energies $\pm \varepsilon$.  The small difference
between the energies of the symmetric and antisymmetric states is
potentially dangerous for the application of the optimal
fluctuation method to the FGM.  The problem arises in the
calculation of the contribution of the disorder realizations
close to the optimal fluctuation:
\[
\eta(x) = {\bar \eta}(x) + \delta \eta(x)\;.
\]
The perturbation $\delta {\hat h} = \sigma_1 \delta \eta(x)$ can,
in principle, strongly mix the symmetric and antisymmetric
intragap states, because of the small energy denominator $2
\varepsilon$ appearing in the perturbation series.  Such mixing
would affect the value of the energy splitting between the two
states.  This problem arises because of the charge conjugation
symmetry of the Dirac Hamiltonian Eq.(\ref{Dirac}), which implies
the symmetry of the spectrum of its eigenvalues around
$\varepsilon = 0$.  We address this question in Appendix A, where
we show that, just because of the symmetry of the Hamiltonian
Eq.(\ref{Dirac}), the mixing of the symmetric and antisymmetric
states is small and thus, despite the small energy splitting, the
optimal fluctuation method is applicable.  The result of Appendix
A also justifies our calculation of the optical absorption
coefficient presented in Ref.~\CITE{MKPL}.

\section{Functional integration}
\label{funint}

The weight of the optimal fluctuation, given by
Eq.(\ref{weight}), is the main factor that describes the
suppression of the density of states at small energy.  To obtain
a full expression for the density, however, one has to take into
account the contribution of the disorder realizations close to
the optimal configuration.  In this section we perform this
calculation using the correspondence between the averaging over
disorder realizations $\eta(x)$ and the quantum-mechanical
averaging over the ground state for a certain double-well
potential.  Then the main suppression factor Eq.(\ref{weight}),
as well as the correction to it, can be easily found using
well-known methods developed for the double-well problem.

\subsection{The corresponding quantum-mechanical problem}
\label{qmprob}

The density of electron states averaged over the white-noise
disorder can be written in the form of the functional integral:
\begin{equation}
\label{fun}
\rho (\varepsilon)  =  \frac{1}{L}
\int\!\!D\eta(x)
\exp \left( -\frac{S}{A} \right)
\mbox {Tr} \left[ \delta ({\hat h} - \varepsilon) \right] \;,
\end{equation}
with
\begin{equation}
\label{S}
S = \frac{1}{2}\int_{-L/2}^{L/2}\!\!dx
\eta^2(x)\;,
\end{equation}
and ${\hat h}$ as defined in Eq.(\ref{Dirac}).  The relation
between the functional integral and a particular quantum
mechanical problem will become apparent after we make a
substitution of variables that allows us to express the disorder
fluctuation $\eta(x)$ in terms of the eigenfunction $\psi(x)$ of
Eq.(\ref{Dirac}) with energy $\varepsilon$. Due to time-reversal
symmetry of the Dirac Hamiltonian its eigenfunctions $\psi(x)$
can be chosen to satisfy
\begin{equation}
\sigma_1 \psi(x) = \psi(x)\;,
\label{timerev}
\end{equation}
and, therefore, can be written in the form:
\begin{equation}
\psi = \left(
\begin{array}{l} u \\ u^{\ast} \end{array} \right) \;.
\end{equation}
The two real variables $w$ and $\phi$ introduced by
\begin{equation}
u = w \left( \cosh(\phi) + i \sinh(\phi) \right)\;,
\end{equation}
then satisfy,
\begin{equation}
\label{sub}
- v_F \frac{d\phi}{dx} + \varepsilon \cosh(2\phi(x))
- \Delta_0 = \eta(x)\;,
\end{equation}
and
\begin{equation} v_F
\frac{d\ln w(x)}{dx} = - \varepsilon \sinh(2\phi(x)) \;.
\end{equation}

\begin{figure}[htbp]
\centering \leavevmode
\epsfxsize=7cm \epsfbox{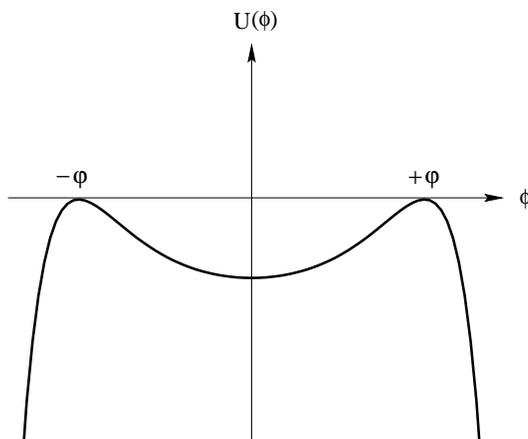}
\caption{
\footnotesize
The inverted double-well potential $U(\phi)$. The instanton
trajectory starts at the left maximum $-\varphi$ and ends at the
right maximum $+\varphi$.
\label{assfig}}
\end{figure}

We now use Eq.(\ref{sub}) to express the disorder
configuration $\eta(x)$ in terms of $\phi(x)$.  Then $S$,
defined by Eq.(\ref{S}), can be written as
\begin{equation}
\label{ACTION}
S[\phi(x)] = S_E[\phi(x)] + S_B\;,
\end{equation}
where
\begin{equation}
\label{eaction}
S_E[\phi(x)] = \frac{1}{2} \int_{-L/2}^{L/2}\!\!dx
\left( v_F^2 \left(
\frac{d\phi}{dx} \right)^2 + \left(\Delta_0 -
\varepsilon \cosh(2 \phi) \right)^2 \right)\;,
\end{equation}
and
\begin{equation}
\label{baction}
S_B = \frac{1}{2} \left.  v_F \left( 2 \Delta_0
\phi - \varepsilon \sinh (2 \phi) \right)
\right|^{+L/2}_{-L/2}\;.
\end{equation}
$S_E[\phi(x)]$ has the
form of the Euclidean action of a particle with mass $v_F^2$ and
coordinate $\phi$ moving in the imaginary time $x$ in the
inverted double-well potential,
\begin{equation}
U(\phi) = - \frac{\Delta_0^2}{2} \left( 1 -
\frac{\cosh(2\phi)}{\cosh(2\varphi)} \right)^2 \;,
\label{pot}
\end{equation}
plotted in Fig.~\ref{assfig}. The angle
$\varphi$ in Eq.(\ref{pot}) is defined by
\begin{equation}
\label{varphi}
\cosh(2\varphi) = \frac{\Delta_0}{\varepsilon}\;.
\end{equation}

Finally, in the functional integral Eq.(\ref{fun}), the disorder
strength $A$ plays the role of $\hbar$.  Thus, we have identified
the quantum double-well problem that corresponds to our disorder
average. A similar correspondence exists also for the random
potential problem.\cite{Igor}

\subsection{The instanton solution}
\label{instsol}

The instanton ${\bar \phi(x)}$ is the solution of the classical
equation of motion in the inverted double-well potential with the
boundary conditions
\begin{equation}
\label{instbc}
{\bar \phi}(\pm \infty) = \pm \varphi\;.
\end{equation}
This corresponds to infinitely long motion from the left maximum
to the right one.  The energy on this classical trajectory,
\begin{equation}
\label{energy}
E = \frac{v_F^2}{2} \left( \frac{d{\bar \phi}}{dx} \right)^2 -
U({\bar \phi}) \;,
\end{equation}
equals zero, which gives,
\begin{equation}
\label{difur}
v_F \frac{d{\bar \phi}}{dx} =
\pm \left( \Delta_0  -
\varepsilon \cosh(2{\bar \phi}(x))  \right)\;.
\end{equation}
As the solution with the minus sign corresponds to the trivial
disorder configuration $\eta(x) = 0$ [cf.~Eq.(\ref{sub})], we
choose the plus sign.  Equation (\ref{difur}) is easily solved,
yielding
\begin{equation}
\label{solution}
e^{2 {\bar \phi(x)}} = \frac
{\cosh \left( K \left( x - x_0 + \frac{R}{2} \right) \right)}
{\cosh \left( K \left( x - x_0 - \frac{R}{2} \right) \right)}
\;,
\end{equation}
where $K$ and $R$ are defined by
\begin{equation}
\label{K}
K = \frac{\Delta_0}{v_F} \tanh (2 \varphi) \;,
\end{equation}
and
\begin{equation}
\label{KR}
K R = 2 \varphi \;.
\end{equation}

Using Eq.(\ref{sub}) we find that the instanton
Eq.(\ref{solution}) is exactly the optimal disorder
configuration given by Eq.(\ref{Soan}) and that
Eqs.~(\ref{varphi}), (\ref{K}), and (\ref{KR}) defining the
parameters of the instanton solution coincide with
Eqs.~(\ref{VFK1}) and (\ref{EPS}).  Finally, for the instanton
trajectory, the value of $S$ defined by Eq.(\ref{ACTION}) is:
\begin{equation}
\frac{S_0}{A} = \frac{S[{\bar \phi(x)}]}{A} =
\frac{2}{g} \left( 2\varphi - \tanh(2\varphi) \right) \;.
\label{Iaction}
\end{equation}

\subsection{Boundary conditions}
\label{bouncond}

We note that the boundary conditions Eq.(\ref{instbc}) for the
instanton trajectory are not, in general, the boundary conditions
that one usually imposes on the eigenfunction $\psi(x)$.  In
particular, $\varphi$ defined by Eq.(\ref{varphi}), depends on
the energy $\varepsilon$, while, for instance, the boundary
conditions for a finite Peierls chain with an even number of
atoms in the continuum limit take the form \cite{Kiv}
\begin{equation}
\frac{\mbox {Im} \left( u(\pm L/2) \right)}
{\mbox {Re} \left( u(\pm L/2) \right)}
= \pm 1 \;,
\end{equation}
which is equivalent to
\begin{equation}
\label{inf}
\phi_{R,L} = \left. \phi \right|_{\pm L/2} = \pm \infty \;.
\end{equation}
We discuss the instanton for these boundary conditions in
Appendix B.  It is clear, however, that in the limit $L
\rightarrow \infty$ the density of states should not depend on
the boundary conditions.  Therefore, we choose here the ones that
lead to the easiest calculation of the density of states, namely,
\begin{equation}
\label{bc}
\phi_{R,L} = {\bar \phi} \left( \pm \frac{L}{2} \right)\;,
\end{equation}
where ${\bar \phi}(x)$ is the instanton solution given by
Eq.(\ref{solution}) with $x_0 = 0$. With the last condition
${\bar \phi}(x)$ is antisymmetric, so that
\begin{equation}
\phi_L = - \phi_R\;.
\end{equation}

\subsection{Regularization}
\label{regularization}

In order to perform the functional integration in Eq.(\ref{fun}),
we first have to regularize it.  We do this by dividing the range
of variation of $x$, $[-\frac{L}{2},\frac{L}{2}]$, into $N$ small
intervals of size $a$, assuming the value of $\eta$ to be
constant inside each interval.  Then,
\begin{equation}
\frac{1}{2A}\int_{-L/2}^{L/2}\!\!dx \eta^2(x)  =
\frac{a}{2A}\sum_{n=1}^{N} \eta_n^2 \;,
\end{equation}
and the integration measure,
\begin{equation}
D\eta(x) = \prod_{n=1}^{N}
\frac{d\eta_n}{\left[\frac{2 \pi A}{a}\right]^{\frac{1}{2}}}\;,
\end{equation}
is defined in such a way that the average of the unity operator
is one.

In the regularized version of the substitution Eq.(\ref{sub}),
the new variables $\phi_n$ ($n = 1,\ldots,N$) are the values of
$\phi$ on the right end of each interval and
\begin{equation}
\phi_0 = \phi_L\;,
\end{equation}
is fixed by the boundary condition at the chain's left end.  For
the moment we treat Eq.(\ref{sub}) as a functional substitution
and forget about the boundary condition at the right end.  Up to
second order in $a$, the regularized form of the substitution
reads:
\begin{equation}
\label{regsub}
v_F \frac{(\phi_n - \phi_{n-1})}{a}  =
\frac
{\left( \varepsilon \cosh(2\phi_{n-1}) - \Delta_0 - \eta_n \right)}
{\left( 1 - \frac{a \varepsilon}{v_F} \sinh(2\phi_{n-1}) \right)}\;\;.
\end{equation}
It is necessary to retain the higher order term in powers of $a$
for the correct calculation of the Jacobian of the substitution.
The integration measure can now be written as
\begin{equation}
D\eta(x) = \prod_{n=1}^{N}
\frac{v_F d\phi_n}{\left[2 \pi a A\right]^{\frac{1}{2}}} J\;,
\end{equation}
with
\begin{equation}
J = \prod_{n=1}^{N}
\left( 1 - \frac{a \varepsilon}{v_F} \sinh(2 \phi_{n-1}) \right)
\approx \exp \left( - \frac{\varepsilon}{v_F}
\int_{-L/2}^{L/2}\!\!dx \sinh (2 \phi(x))
\right) \;.
\label{J}
\end{equation}

It is important that Eq.(\ref{regsub}) provides a one-to-one
correspondence beween the disorder configuration $\{\eta_n\}$ and
the values of the solution of the differential equation
(\ref{sub}) at the points $x = a n\;,\;n = 1,\ldots,N$.  In other
words, Eq.(\ref{regsub}) relates the disorder configuration
to the electron wave function in the presence of this disorder,
which gives us a convenient representation for the trace of the
delta-dunction in Eq.(\ref{fun}).  The energy $\varepsilon$ is
the eigenvalue of the Hamiltonian ${\hat h}$ if $\phi$ at the
right end satisfies the boundary condition:
\begin{equation}
\phi_N = \phi_R\;.
\end{equation}
Therefore, the average density of states Eq.(\ref{fun}) can
be written in the form:
\begin{equation}
\rho (\varepsilon)  =  \frac{1}{L}
\int\!\!D\eta(x)
\exp \left( -\frac{S}{A} \right)
\left( \frac{\partial \phi_N}{\partial \varepsilon}
\right)_{\phi_0,\eta(x)}
\delta\left( \phi_N - \phi_R \right)  \;.
\end{equation}
The delta-function removes the integration over $\phi_N$ and,
using Eq.(\ref{sub}), one can easily calculate the derivative of
$\phi_N$ with respect to $\varepsilon$, keeping both the disorder
$\eta(x)$ and the value of $\phi$ at the left end of the chain
fixed:
\begin{equation}
\label{pardir}
\left( \frac{\partial \phi_N}{\partial \varepsilon}
\right)_{\phi_0,\eta(x)} =
\frac{y(\frac{L}{2})}{v_F} \int_{-L/2}^{L/2}\!\!dx
\frac{\cosh(2 \phi(x))}{y(x)} \;.
\end{equation}
Here, $y(x)$ is given by
\begin{equation}
y(x) = \exp \left(
\frac{2 \varepsilon}{v_F}
\int_{-L/2}^{x}\!\!dx^{\prime}
\sinh(2 \phi(x^{\prime})) \right)\;,
\label{y(x)}
\end{equation}
so that $J$, defined by Eq.(\ref{J}), can be written as
\begin{equation}
\label{Jac}
J = \frac{1}{\sqrt{y \left( \frac{L}{2} \right)}}\;.
\end{equation}

We thus arrive at the following final form of the functional
integral
\begin{equation}
\label{fun1}
\rho (\varepsilon)  = \frac{1}{L}
\int\!\!D^{\prime}\phi(x)
\exp \left( -\frac{S_E}{A} \right)
F[\phi(x)]\;,
\end{equation}
where the integration measure is defined by
\begin{equation}
D^{\prime}\phi(x) =
\frac{v_F}{\left[2 \pi a A\right]^{\frac{1}{2}}}
\prod_{n=1}^{N - 1}
\frac{v_F d\phi_n}{\left[2 \pi a A\right]^{\frac{1}{2}}} \;,
\end{equation}
and
\begin{equation}
F[\phi(x)] = J
\left( \frac{\partial \phi_N}{\partial \varepsilon}
\right)_{\phi_0,\eta(x)}
\exp \left( -\frac{S_B}{A} \right)\;.
\end{equation}

\subsection{Saddle-point integration}
\label{spintegration}

In this section we perform the functional integration in
Eq.(\ref{fun1}) over configurations $\phi(x)$ in the vicinity
of the saddle-point configuration $\phi_\ti{sp}(x)$ defined as
\begin{equation}
\left. \frac{\delta S}{\delta \phi(x)} \right|_{\phi_\ti{sp}(x)}
= 0\;.
\end{equation}
The boundary conditions Eq.(\ref{bc}) were chosen to ensure
that $\phi_\ti{sp}(x) = {\bar \phi}(x)$, where ${\bar
\phi}(x)$ is given by Eq.(\ref{solution}) with $x_0 = 0$.

The result of the integration is:
\begin{equation}
\label{sadpoint}
\rho (\varepsilon) = \frac{J D}{L}
\left( \frac{\partial \phi_N}{\partial \varepsilon}
\right)_{\phi_0,{\bar \eta(x)}}
\exp \left( - \frac{S_0}{A} \right)\;,
\end{equation}
where $S_0$ is given by Eq.(\ref{Iaction}) and $D$ is the
determinant of the operator
\begin{equation}
-\frac{v_F^2}{2} \frac{d^2}{dx^2} +
4 \varepsilon
\left( \varepsilon \cosh(4 \phi(x))
- \Delta_0 \cosh(2 \phi(x)) \right)\;,
\end{equation}
obtained by the second variation of the Euclidean action
Eq.(\ref{eaction}).  In Eq.(\ref{sadpoint}) $J$, $D$, and
$\partial \phi_N / \partial \varepsilon$ have to be
calculated at $\phi(x) = {\bar \phi}(x)$.

Using Eq.(\ref{difur}) (with the plus sign) one easily finds
that
\begin{equation}
\frac{2 \varepsilon}{v_F} \sinh(2{\bar \phi}) =
- \frac{d}{dx} \ln \left( \frac{d{\bar \phi}}{dx} \right)\;,
\end{equation}
so that Eq.(\ref{y(x)}) can be written in the form
\begin{equation}
y(x) = \frac{\frac{d{\bar \phi}}{dx} \left( -\frac{L}{2} \right)}
{\frac{d{\bar \phi}}{dx} ( x )}\;.
\end{equation}
Since ${\bar \phi}(x)$ is antisymmetric, $y\left( \frac{L}{2}
\right) = 1$, so that $J = 1$ and
\begin{equation}
\left( \frac{\partial \phi_N}{\partial \varepsilon}
\right)_{\phi_0,{\bar \eta}(x)} = \frac
{\sinh\left( 2{\bar \phi}\left( \frac{L}{2} \right) \right)}
{v_F \frac{d{\bar \phi}}{dx} \left( \frac{L}{2} \right)}\;.
\end{equation}
As for large $L$, $\sinh\left( 2{\bar \phi}(\frac{L}{2}) \right)$
can be replaced by $\sinh(2 \varphi)$ and
\begin{equation}
\frac{d{\bar \phi}}{dx} \left( \frac{L}{2} \right) \approx
2 K \sinh(2\varphi)\;e^{-KL} \;,
\end{equation}
we obtain
\begin{equation}
\label{pd}
\left( \frac{\partial \phi_N}{\partial \varepsilon}
\right)_{\phi_0,{\bar \eta(x)}} = \frac{e^{KL}}{2 v_F K} \;.
\end{equation}

We calculate the determinant $D$ using the standard
method for calculation of determinants of Schr\"odinger
operators in one dimension, which can be found in
Ref.~\CITE{Coleman}.  The result is
\begin{equation}
\label{det}
D = L \frac{(2 K v_F)^2}{\pi A} \sinh(2\varphi)\;e^{-KL} \;.
\end{equation}
The determinant is proportional to the chain size, because the
soliton-antisoliton pair can appear at any place in the chain
(formally the factor $L$ comes from the integration over the zero
mode \cite{Coleman}).  Taking all factors together, we
finally obtain for the averaged density of states
\begin{equation}
\rho (\varepsilon)  =
\frac{2 K v_F}{ \pi A} \sinh(2\varphi)\;
\exp \left( - \frac{S_0}{A} \right) \;.
\end{equation}
The resulting expression for the average density of electron
states in the limit $\varepsilon \ll \Delta_0$ is
\begin{equation}
\rho (\varepsilon)  =
\frac{e}{\pi g v_F}
\left( \frac{e \varepsilon}{2 \Delta_0}
\right)^{\frac{2}{g} - 1}\;.
\label{ansdos}
\end{equation}
For $g \ll 1$, this agrees with Eq.(\ref{dos}), confirming the
validity of the optimal fluctuation method at small energies
and weak disorder.

\section{Discussion}
\label{discussion}

In this paper, we have used functional integration to establish
a relation between the Fluctuating Gap Model, which describes
one-dimensional electron motion in the presence of both a Peierls
distortion and quenched disorder, and the quantum motion of a
particle in a double-well potential.  Averaging over disorder
realizations in the model with quenched disorder corresponds to
the sum over all paths in the quantum-mechanical problem.  We
have shown that the instanton trajectory describing the
imaginary-time motion between the two wells corresponds to the
optimal disorder fluctuation.  The probability of this
fluctuation determines the asymptotic behavior of
the averaged electron density of states.

We showed that the most probable form of the wave function of the
electron states lying deep within the pseudogap contains two
peaks.  The optimal disorder fluctuation that induces
such a state, has the form of a soliton-antisoliton pair and the
peaks of the wave function are localized near the two kinks of
this fluctuation (see Fig.~\ref{s-a}).  As we demonstrated in
this paper and in Ref.~\CITE{MKPL}, the instanton approach
allows for a relatively easy calculation of the density of states
and absorption coefficient.  Our result Eq.(\ref{ansdos}) is
valid if the density of disorder-induced states is small, which
is the case when $|\varepsilon| \ll \Delta_0$ and $g \ll 1$.

Above, we emphasized the strong relation between the calculation
of the disorder average and the double-well problem.  However,
there is also a clear difference: In the double-well problem, an
important role is played by multi-instanton configurations, in
which instantons are followed by anti-instantons and {\em vice
versa}.\cite{Coleman} Although the action for the configuration
containing $n$ instantons and anti-instantons is $n$ times the
action of a single instanton $S_\ti{inst}$, the multi-instanton
contribution obtains a factor $T^n / n!$ from the sum over all
possible instanton positions ($T$ is the imaginary time
corresponding to $L$).  This statistical factor grows with $T$,
while the suppression factor $\exp(- n S_\ti{inst} / \hbar)$ does
not decrease any further, making the multi-instanton
configurations important at large enough $T$.  We note that it is
the sum over all possible numbers of instantons and
anti-instantons that gives the energy splitting between the two
lowest states in the double-well potential.\cite{Coleman}

By contrast, the asymptotic expression Eq.(\ref{ansdos}) for the
density of states in the FGM is given by only one instanton,
which means that the largest contribution to the averaged density
of states comes from a single disorder fluctuation.  The
origin of the difference lies in the fact that, while in the
quantum double-well problem the anti-instanton describes a real
physical process (the tunneling between the wells in the
direction opposite to the one described by the instanton), in the
FGM it corresponds to zero disorder.  To see this, we note that
the instanton $\phi_\ti{a}(x) = {\bar \phi}(-x)$, obtained from
the instanton by ``time-reversal'', $x \rightarrow - x$, is
the solution of Eq.(\ref{difur}) with the minus sign.  Comparing
this equation with Eq.(\ref{sub}), we find that $\phi_\ti{a}(x)$
corresponds to $\eta(x) =0$.  The Euclidean action $S_E$ of the
anti-instanton is exactly cancelled by the ``boundary'' term
$S_B$ [cf.~Eq.(\ref{ACTION})].  It is then clear, that
anti-instantons (and, hence, the multi-instanton configurations)
cannot play any role for the calculation of the density of
states.  Formally, this happens because for $\phi(x) =
\phi_\ti{a}(x)$ we have: \begin{equation} \label{pda} \left(
\frac{\partial \phi_N}{\partial \varepsilon}
\right)_{\phi_0,{\bar \eta(x)}} \approx \frac{\Delta_0}{v_F K
\varepsilon} + \frac{2 L K}{\varepsilon} e^{-KL}\;,
\end{equation} which tends to a constant value in the large $L$
limit, while the corresponding expression for the instanton
solution Eq.(\ref{pd}) diverges as $\exp(KL)$.  Since the value
of the determinant $D \propto L \exp(-KL)$ for the anti-instanton
is the same as for the instanton, the anti-instanton contribution
is proportional to $L \exp(-KL)$ and, therefore, vanishes in the
$L \rightarrow \infty$ limit.

\section*{Acknowledgements}

The authors are grateful to Dr. I.  V.  Kolokolov for useful
discussions.  This work is supported by the "Stichting
Scheikundig Onderzoek in Nederland (SON)" and the "Stichting voor
Fundamenteel Onderzoek der Materie (FOM)".

\appendix

\section{The role of symmetries of the FGM Hamiltonian}
\label{appendixA}

In this appendix we show that owing to the symmetry properties of
the FGM Hamiltonian a small perturbation of the optimal
fluctuation
\begin{equation}
\eta(x) = {\bar \eta}(x) + \delta \eta(x)\;,
\label{perturbation}
\end{equation}
does not strongly mix the symmetric and antisymmetric states, despite
the small energy splitting between them. As was explained in
Sec.~\ref{FGM} such a mixing, if it would occur, would
invalidate our optimal fluctuation calculation.

First, we obtain the effective Hamiltonian, acting on the
subspace of the two intragap states, which includes the
virtual excitations to all other electron states.  To this
end we write the single-electron Hamiltonian for the
disorder realization $\eta(x)$, given by
Eq.(\ref{perturbation}), in the form:
\begin{equation}
{\hat h} = {\hat h}^{(0)} + {\hat h}^{(1)}\;,
\end{equation}
where
\begin{equation}
{\hat h}^{(0)}  =
\sigma_3 \frac{v_F}{i} \frac{d}{dx}
+ \sigma_1 \left( \Delta_0 + {\bar \eta}(x) \right)\;,
\end{equation}
and ${\hat h}^{(1)}$ is the perturbation:
\begin{equation}
{\hat h}^{(1)}  = \sigma_1  \delta \eta(x)  \;.
\end{equation}

We divide the Hilbert space of the Hamiltonian ${\hat
h}^{(0)}$ into two subspaces: A and B.  The subspace A consists
of the symmetric and antisymmetric intragap states, which here
are denoted as $|\pm\rangle$, while all the other states
belong to the subspace B, so that
\begin{equation}
{\hat P}_A + {\hat P}_B = 1\;,
\end{equation}
where ${\hat P}_{A,B}$ are the projection operators for the
corresponding subspaces.  The effective energy-dependent
Hamiltonian, acting on the subspace A, has the form:
\begin{equation}
{\hat h}_A(\varepsilon) = {\hat h}_{A}^{(0)} +
{\hat h}_{A}^{(1)} + {\hat h}_{A}^\ti{(v)}(\varepsilon)
\label{h_A}
\end{equation}
where the first term is the diagonal $2$-by-$2$ matrix:
\begin{equation}
{\hat h}_{A}^{(0)} = {\hat P}_A {\hat h}^{(0)} {\hat P}_A =
\left( \begin{array}{cc}
+ \varepsilon_0(R) & 0 \\
0 & - \varepsilon_0(R)
\end{array}
\right)\;,
\end{equation}
the second term is the first-order perturbation,
\begin{equation}
{\hat h}_{A}^{(1)} = {\hat P}_A {\hat h}^{(1)} {\hat P}_A\;,
\end{equation}
and the last term,
\begin{equation}
{\hat h}_{A}^\ti{(v)}(\varepsilon) =
{\hat P}_A {\hat h}^{(1)}
\frac{1}{\varepsilon - {\hat h}^{(0)} - {\hat P}_B {\hat h}^{(1)}}
{\hat P}_B {\hat h}^{(1)} {\hat P}_A\;,
\label{virtex}
\end{equation}
describes the higher-order processes involving virtual
excitations to the subspace B.

We now want to show that the non-diagonal matrix
elements of the effective Hamiltonian ${\hat h}_A(\varepsilon)$
are smaller than its diagonal matrix elements.
First, we show that the first-order nondiagonal matrix elements
equal $0$:  On the one hand, using the relation
$\psi_{\pm}^{\ast}(x) = \sigma_1 \psi_{\pm}(x)$ [cf.
Eq.(\ref{timerev})] we obtain that the first-order
perturbation Hamiltonian is real and symmetric:
\begin{equation}
\langle + | {\hat h}^{(1)} | - \rangle  =
\langle + | \sigma_1 {\hat h}^{(1)} \sigma_1 | - \rangle =
\langle + | {\hat h}^{(1)} | - \rangle^{\ast}  =
\langle - | {\hat h}^{(1)} | + \rangle \;.
\label{hand1}
\end{equation}
On the other hand, we can use the relation between the
positive-energy and negative-energy eigenstates of the Dirac
Hamiltonian, $\psi_{\pm}(x) = - \sigma_2 \psi_{\mp}(x)$,
to show that
\begin{equation}
\langle + | {\hat h}^{(1)} | - \rangle  =
- \langle + | \sigma_2 {\hat h}^{(1)} \sigma_2 | - \rangle =
- \langle - | {\hat h}^{(1)} | + \rangle \;.
\label{hand2}
\end{equation}
Comparing Eq.(\ref{hand1}) with Eq.(\ref{hand2}) we obtain:
\begin{equation}
\langle + | {\hat h}^{(1)} | - \rangle =
\langle - | {\hat h}^{(1)} | + \rangle = 0\;.
\label{ndzero}
\end{equation}

Next we consider the non-diagonal matrix elements of ${\hat
h}_{A}^\ti{(v)}(\varepsilon)$.  The perturbative expansion of
${\hat h}_{A}^\ti{(v)}(\varepsilon)$ in powers of ${\hat
h}^{(1)}$ describes virtual excitations into the subspace B, so
that all intermediate states in this expansion are separated from
the intragap states by a large energy ($\gg \varepsilon_0(R)$).
Since $|\varepsilon| \sim \varepsilon_0(R)$, the expansion terms
involved do not contain small energy denominators.  Moreover, the
non-diagonal matrix elements of the operator (\ref{virtex}) equal
$0$ at $\varepsilon = 0$.  The proof of this statement is
analogous to the proof of Eq.(\ref{ndzero}) and we shall not
repeat it here.\footnote{The important point is that in the basis
of the eigenstates of ${\hat h}^{(0)}$, satisfying
Eq.(\ref{timerev}), ${\hat h}^{(0)}$ and ${\hat h}^{(1)}$ are
real symmetric matrices and that they both anticommute with
$\sigma_2$.} Hence, the non-diagonal matrix elements of the
operator ${\hat h}_{A}^\ti{(v)}(\varepsilon)$ are
$O(\varepsilon)$.  Moreover, they contain an additional small
factor $\delta\eta(x) / \Delta_0$.  Therefore, the small disorder
perturbation $\delta\eta(x)$ cannot result in a strong mixing of
the two intragap states.

Next we show that the corrections to the energies of these two
states due to the diagonal matrix elements of the effective
Hamiltonian are also small, which will justify the validity of
our calculation of the optical absorption coefficient.\cite{MKPL}
Using the expressions for the wave functions of the symmetric and
antisymmetric states,\cite{MKPL} the straightforward calculation of
the first-order terms gives
\begin{equation}
\langle + | {\hat h}^{(1)} | + \rangle =  -
\langle - | {\hat h}^{(1)} | - \rangle =
\int\!\!dx \delta\eta(x) \phi(x - \frac{R}{2})
\phi(x + \frac{R}{2})\;,
\end{equation}
where
\begin{equation}
\phi(x) = \sqrt{\frac{K}{2}} \frac{1}{\cosh(K x)}.
\end{equation}
Since the condition $\varepsilon_0(R) \ll \Delta_0$ implies $R
\gg \xi_0$ [see Eq.(\ref{dist})], the last equation can be
written in the form:
\begin{equation}
\langle + | {\hat h}^{(1)} | + \rangle \approx
\frac{2}{\xi_0} e^{-\frac{R}{\xi_0}}
\int_{-R/2}^{R/2}\!\!dx \delta\eta(x) \approx
\varepsilon_0(R) \int_{-R/2}^{R/2}\frac{dx}{\xi_0}
\frac{\delta\eta(x)}{\Delta_0}\;,
\end{equation}
{\em i.e.}, the first-order correction to the energy of the
intragap state with the energy $\pm \varepsilon_0(R)$ is
proportional to this small energy times a factor
$O\left(\frac{\delta \eta(x)}{\Delta_0}\right)$.  This result can
be easily traced back to the fact that the energy of the midgap
state of a single kink cannot be changed by a small perturbation
of $\Delta(x)$ (see, {\em e.g.}, Ref.~\CITE{Witten}) and,
therefore, it holds in all orders of the expansion of the
effective Hamiltonian in powers of $\delta \eta(x)$.

Thus, we found that, despite the small energy splitting between
the symmetric and antisymmetric intragap states, a small
variation of the disorder fluctuation does not strongly affect
either the wave functions, or the energies of these two states.
This result is a direct consequence of symmetries of the
Hamiltonian Eq.(\ref{Dirac}).

\section{Different boundary conditions}
\label{appendixB}

In this appendix we discuss the instanton trajectory for the
boundary conditions Eq.(\ref{inf}), which one obtains in the
continuum limit for an open chain with an even number of atoms.
To avoid spurious divergences related to an infinite value of
$\phi$ at the boundaries, we will use the following procedure.
First, we impose, instead of Eq.(\ref{inf}), the boundary
conditions,
\begin{equation}
\label{ninf}
\phi_{R,L} = \left. \phi \right|_{\pm L/2} = \pm \Phi \;\;,
\end{equation}
where $\Phi \gg \varphi$ is large but finite. Next we take the
limit $L \rightarrow \infty$ and only then $\Phi$ is
tended to infinity.

Since for $|\Phi| > \phi$ part of the instanton trajectory ${\bar
\phi(x)}$ lies outside the well of the potential Eq.(\ref{pot})
located between $-\varphi$ and $\varphi$, the energy $E$ defined
by Eq.(\ref{energy}) is positive.  For $E > 0$ the equation for
the instanton trajectory is,
\begin{equation}
\label{difur1}
v_F \frac{d{\bar \phi}}{dx} =
\sqrt{
\left( \Delta_0  - \varepsilon \cosh(2{\bar \phi}(x))  \right)^2
+ 2E}\;\;.
\end{equation}

Now, the particle's velocity becomes very large as soon as it
leaves the well. The large Euclidean action Eq.(\ref{eaction})
associated with this region is, however, completely compensated
by the surface term Eq.(\ref{baction}) in the infinite $L$ limit.
To see this, one notes that even for infinite $\Phi$ the time of
motion $L$ along the instanton trajectory is finite, as the
velocity grows very fast outside the well.  We are, however,
interested in long time (large $L$) trajectories.  The only way
to make $L$ large is to tend the positive energy $E$ to zero, in
which case the particle would spend most of the time near the
maxima $-\varphi$ and $\varphi$ of the potential $U(\phi)$, where
it has a small velocity.  The relation between the energy and the
time of motion for large $L$ is,
\begin{equation}
E \approx 8 \frac{\left(K v_F \right)^4}{\varepsilon^2}
e^{-2 \varphi} e^{- K L}\;\;.
\end{equation}
For $\Phi \gg \varphi$ the dependence of the energy on $\Phi$ can
be neglected.

When the energy $E$ tends to zero, the trajectory given by
Eq.(\ref{difur1}) inside the well, {\em i.e.}, for $|\phi| <
\varphi$, tends to the solution of Eq.(\ref{difur}) with the plus
sign.  In other words, in the limit $L \rightarrow \infty$, this
part of the trajectory becomes the instanton solution
Eq.(\ref{solution}) that we considered above.  On the other hand,
outside the well ($|\phi| > \varphi$) the solution of
Eq.(\ref{difur1}) tends to the solution of Eq.(\ref{difur}) with
the minus sign.  The latter corresponds to zero disorder,
$\eta(x) = 0$, which explains the cancellation between the
Euclidean action coming from outside the well and the surface
term.  We proved, therefore, that in the limit $L \rightarrow
\infty$ the suppression factor $S$ for the instanton with
boundary conditions Eq.(\ref{ninf}) is identical to
Eq.(\ref{Iaction}).

The boundary conditions Eq.(\ref{ninf}) imply that the solution
${\bar \phi}(x)$ is antisymmetric.  Therefore, the Jacobian $J$
given by Eq.(\ref{Jac}) equals unity, as before.  At the same
time, however, the values of the partial derivative $\left(
\frac{\partial \phi_N}{\partial \varepsilon}
\right)_{\phi_0,{\bar \eta(x)}}$ and the determinant $D$ differ
considerably from Eqs.~(\ref{pd}) and (\ref{det}), respectively.
For instance,
\begin{equation}
\left( \frac{\partial \phi_N}{\partial
\varepsilon} \right)_{\phi_0,{\bar \eta(x)}} =
\frac{e^{2 \varphi} e^{4 \Phi} e^{K L}}{16 \sinh^4(2
\varphi)}\;\;.
\end{equation}
Nevertheless, the product of the determinant and the partial
derivative remains unchanged up to small corrections that vanish
for $L \rightarrow \infty$.  Therefore, despite the fact that the
instanton trajectories for the boundary conditions Eq.(\ref{inf})
and Eq.(\ref{bc}) are rather different, the average density of
states obtained by the saddle-point integration near these
trajectories is the same for sufficiently long chains, as, of
course, should be the case.

\end{document}